# A Recent Survey of Vision Transformers for Medical Image Segmentation

*Asifullah Khan[1, 2, 3*], Zunaira Rauf[1, 2], Abdul Rehman Khan[1], Saima Rathore[4, 5], Saddam Hussain Khan[6], Najmus Saher Shah[1], Umair Farooq[1], Hifsa Asif[1, 7], Aqsa Asif[1, 7], Umme Zahoora[1], Rafi Ullah Khalil[8], Suleman Qamar[9], Umme Hani Asif[9], Faiza Babar Khan[9], Abdul Majid[9] and Jeonghwan Gwak[10*]*

[1]Pattern Recognition Lab, Department of Computer & Information Sciences, Pakistan Institute of Engineering & Applied Sciences, Nilore, Islamabad 45650, Pakistan
[2]PIEAS Artificial Intelligence Center (PAIC), Pakistan Institute of Engineering & Applied Sciences, Nilore, Islamabad 45650, Pakistan
[3]Center for Mathematical Sciences, Pakistan Institute of Engineering & Applied Sciences, Nilore, Islamabad 45650, Pakistan
[4]Avid Radio pharmaceuticals, Philadelphia, PA, USA
[5]Eli Lilly and Company, Indianapolis, IN, USA.
[6]Department of Computer Systems Engineering, University of Engineering and Applied Sciences (UEAS), Swat, Pakistan
[7]Air University, E-9, Islamabad 44230, Pakistan
[8]Digital Image Processing Lab, Department of Computer Science, Islamia College Peshawar, Pakistan.
[9]Department of Computer and Applied Sciences (DCIS), Pakistan Institute of Engineering & Applied Sciences, Nilore, Islamabad 45650, Pakistan
[10]Department of Software, Korea National University of Transportation, Chungju 27469, Republic of Korea

**Corresponding Authors:** [*]Asifullah Khan, asif@pieas.edu.pk, [*]Jeonghwan Gwak, jgwak@ut.ac.kr.

## Abstract:

Medical image segmentation plays a crucial role in various healthcare applications, enabling accurate diagnosis, treatment planning, and disease monitoring. Traditionally, convolutional neural networks (CNNs) dominated this domain, excelling at local feature extraction. However, their limitations in capturing long-range dependencies across image regions pose challenges for segmenting complex, interconnected structures often encountered in medical data. In recent years, Vision Transformers (ViTs) have emerged as a promising technique for addressing the challenges in medical image segmentation. Their multi-scale attention mechanism enables effective modeling of long-range dependencies between distant structures, crucial for segmenting organs or lesions spanning the image. Additionally, ViTs' ability to discern subtle pattern heterogeneity allows for precise delineation of intricate boundaries and edges, a critical aspect of accurate medical image segmentation. However, they do lack image-related inductive bias and translational invariance, potentially impacting their performance. Recently, researchers have come up with various ViT-based approaches that incorporate CNNs in their architectures, known as Hybrid Vision Transformers (HVTs) to capture local correlation in addition to the global information in the images. This survey paper provides a detailed review of the recent advancements in ViTs and HVTs for medical image segmentation. Along with the categorization of ViT and HVT-based medical image segmentation approaches, we also present a detailed overview of their real-time applications in several medical image modalities. This survey may serve as a valuable resource for researchers, healthcare practitioners, and students in understanding the state-of-the-art approaches for ViT-based medical image segmentation.

## Key words:

# 1. Introduction

In medical image analysis, image segmentation has a significant role [1]. Accurate and efficient segmentation in medical images can help physicians in the identification and evaluation of specific anatomical characteristics, diseases, or regions of interest. These segmented regions are then analyzed to estimate the disease prognosis and to devise effective treatment strategies [2].

Over the past ten years, significant progress has been made in medical image segmentation using deep learning techniques, more especially Convolutional Neural Networks (CNNs) because of their capacity to capture intricate patterns from unprocessed data [3]. CNNs tend to capture local correlation in the images, which allows them to learn the local patterns distributed among the entire image. Their inherent nature to extract 2D features from images gave them an edge over traditional methods. The UNet is among the initial CNN-based encoder-decoder techniques that revolutionized the medical image segmentation tasks [4]. Over time many UNet-like algorithms were proposed for various medical image segmentation tasks, that demonstrated remarkable improvement in various medical image modalities [5]–[7]. However, their capacity to model global relationships in the images is constrained by the local nature of the convolution operator [8].

Researchers have proposed a number of approaches to incorporate global-level information in the CNN-based segmentation architectures [9], [10]. In this regard, various attention-based approaches were developed to focus on the important regions in the medical images [11]. Recently, Vision Transformers (ViTs), which were originally introduced for natural images, have gained prominence in a variety of image-related fields, including computer vision and medical image analysis [12].

The multi-scale self-attention mechanism (MSA) in ViTs possesses the capability to capture global relationships within the images [13]. These ViT-based architectures have demonstrated remarkable results in handling complex medical image segmentation tasks in various image modalities. The inherent ability of ViTs to model global features from medical images involves initially dividing them into patches and subsequently processing each patch through their self-attention mechanism [14]. However, ViT architecture does not include image-related inductive bias to deal with the positional variations in the images. In addition, their performance may be affected by the amount of training data, as ViT models generally require large training data for learning [15].

Recently, there has been a growing interest in integrating CNN inductive biases into ViT architectures, leading to the emergence of CNN-Transformers, also known as Hybrid Vision Transformers (HVTs) [16]. These CNN-Transformer architectures combine the MSA mechanism of ViTs with the convolutional operations of CNNs to model both local and global relationships within images. Several novel frameworks have been developed, leveraging CNN-Transformer architectures and showcasing impressive results in tasks related to medical image segmentation [17].

Due to the rapid surge in medical image segmentation techniques based on ViTs, various interesting surveys have been conducted to review these approaches [7], [11], [15], [18]. Most of these surveys are either focused on a special organ or a special modality. This paper presents a comprehensive survey that spans recent ViT-based segmentation approaches applied to medical images across diverse modalities. In addition, we categorize these methods as 1) ViT-based methods, 2) HVT-based methods. Secondly, we devise a taxonomy for HVT as a) encoder-based integration, b) decoder-based integration, and c) in between encoder-decoder integration.

The goal of this study is to give a thorough overview of the latest techniques, strategies, and developments related to the utilization of ViTs, specifically for medical image segmentation. We intend to provide researchers, clinicians, etc. with the extensive knowledge required to use ViT architectures efficiently for precise and effective medical image segmentation.

The rest of this survey is organized as follows: The basic ideas behind medical image segmentation and ViTs are summarized in Section II and Section III. Section IV provides a thorough analysis of ViT and HVT-based Medical Image Segmentation techniques. We categorize these approaches based on the utilization of ViT and HVT in their encoder-decoder-based architecture. Section V presents a detailed overview of ViT and HVT-based medical image segmentation methods in several image modalities, including CT images, histopathology, MRI, ultrasound, X-Ray, etc. Section VI enlists several techniques utilized by VIT and HVT-based medical image segmentation approaches to enhance their performances. Sections VII and VIII present challenges and future recommendations, and in Section IX the survey paper is concluded.

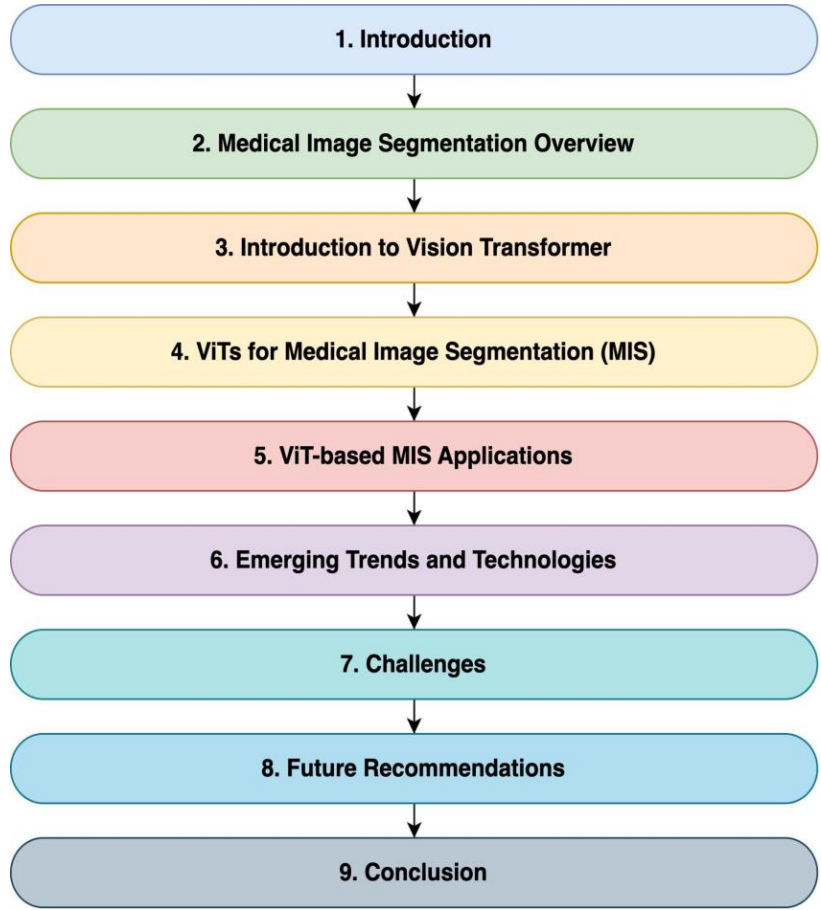

Figure 1: The structure of the survey paper, displaying various sections.

## 2. Medical Image Segmentation Overview

Diagnosis and treatment of disease and other medical conditions require medical tests that help physicians look inside a body. In this regard, medical imaging is essential since it allows for non-invasive viewing and diagnosis of a variety of medical disorders [18]. Different imaging modalities offer unique insights into different aspects of the human body [19].

Medical images are of great importance in diagnosing and treating diseases as they provide valuable information about the symptoms. However, when examining medical images at the pixel level, only certain regions contribute to the accurate diagnosis and treatment, such as the tumor region in a CT scan. Therefore, segmenting the specific areas of interest, whether they are infected regions or abnormal organs, remains a significant and challenging task for researchers [20]. Accurate medical image segmentation is the most widely used task in medical diagnosis and is essential to computer-assisted assessment, image-guided surgery, and therapeutic planning [21]–[23]. Segmentation of an image includes dividing an image into several sections and objects as per the area of interest. Thus, the most vital act of feature segmentation in medical imaging is to precisely identify the region of interest [24], [25].

Segmentation can be divided into the semantic, instance, and panoptic segmentation subtasks. Semantic segmentation is generally regarded as classification at pixel-level i.e. it labels every pixel in an image. Instance segmentation detects and outlines every item or object of concern in an image and also draws bounding boxes around the instances. Whereas, the panoptic segmentation aims to provide a more comprehensive understanding by combining both the semantic and the instance segmentation. It not only assigns a category label (like semantic segmentation) to each pixel in the image but also classifies different instances (like instance segmentation) of objects within the same class.

In clinical pathology, the gold standard is to carry out manual medical image analysis by pathologists. This manual examination involves highly trained pathologists visually examining medical images and manually outlining or marking the regions of interest, such as tumors, lesions, or specific anatomical structures. Manual assessment is tedious, exhausting, and can be influenced by human subjectivity, requiring expert clinicians [26]. However, digital pathology has introduced a new paradigm shift, enabling a fast and accurate diagnosis by providing researchers with the ability to design automated systems. These automated systems can help pathologists by providing them with a fast diagnosis that can be used as a second opinion, thus reducing their burden [27]. Different modalities in medical imaging depict the internal structures of the human body.

Segmentation applied to these modalities can isolate specific organs as well as aid in disease identification within a specific organ.

For automated diagnosis, several traditional methods have been applied for the segmentation of medical images over the years. These methods encompass a range of techniques, including region-based segmentation techniques, edge detection techniques, statistical shape models, active contour models, thresholding, clustering, and machine learning methods [28]. However, automatically segmenting medical images with precision greatly enhances clinical assessment, but remains a formidable challenge due to (1) the wide range of sizes found within objects in medical images, (2) the ambiguity in structure outlines, coupled with their diverse textural variations and complex shapes, which can easily yield inaccurate results, (3) the challenge posed by low-intensity contrast, when isolating objects of interest from the background, and (4) unavailability of sufficient training datasets [22], [23].

In the last few years, the progress in the realm of deep learning, following the triumph of CNNs, has aided significant improvement in the algorithms of medical image segmentation [29]. Deep learning-based segmentation algorithms demonstrate exceptional performance when provided with dense labels, specifically annotations at each pixel. CNNs are effectively applied to medical image segmentation and classification for assessing diverse imaging techniques in clinical practices including ultrasound, MRI, X-ray, etc. Nevertheless, owing to the restricted receptive field of convolution filters, they are unable to capture prolonged dependencies within the medical images [30]. U-Net has been a predominant choice in the field of medical imaging owing to its efficient performance and outstanding accuracy, facilitated by the inclusion of skip connections and the encoder-decoder network up till now [31]. V-Net is an eminent architectural variant of U-Net based on FCN but its operations are in 3D to work with volumetric images [32].

With the advent of ViTs, Transformers have emerged as a viable approach for performing various tasks related to images, including image segmentation, image recognition, and object detection [33]. Transformers, with their capability to learn global context, excel in achieving precise medical image segmentation. This leads to significant performance improvements, especially in organs with large receptive fields, as seen in the segmentation of lungs. Transformers extract global and long-term dependencies because of their self-attention mechanism, whereas CNNs, with their local receptive field, typically learn local correlation in the images. Transformer-based methods have proven highly effective in various real-world segmentation tasks, streamlining the segmentation process across different areas, including brain tumor/tissue segmentation [34]–[36], cell nuclei segmentation [37], colon cancer segmentation [38], liver lesion segmentation [39], kidney tumor segmentation [40], multiple-organ segmentation [41], and skin lesion segmentation [42], [43]. Recently, researchers have combined both architectures to merge their benefits. These advancements have facilitated significant progress in the field of medical image segmentation [44].

## 3. Introduction to Vision Transformers

Over the past decade, CNNs have been extensively utilized for a variety of computer vision tasks, such as the analysis of medical images. However, the limited receptive field of convolution filters restricts CNNs from capturing prolonged dependencies within medical images [45], [46] which may affect their performance. Recently, Transformers and ViTs have gained the predominance attention of researchers due to their outstanding performance. Transformers, which were initially proposed for machine translation, became popular in many natural language processing (NLP) tasks due to their ability to process sequences by utilizing the self-attention mechanism.

Inspired by the success of Transformers in NLP, Dosovitskiy, et al. proposed a ViT for processing images as a sequence of tokens rather than a 2D grid of pixels [13]. The remarkable success of ViT architectures in several image-related tasks encouraged researchers to utilize ViTs with several improvements[47], [48]. These include either changing position encoding techniques, self-attention mechanisms, or creating new architectural variants.

## 3.1. Fundamentals of ViTs

ViTs are based on the Transformer architecture to handle image data, revolutionizing the area of computer vision. By converting the input image pixels into sequences and using a self-attention mechanism to identify long-range dependencies and interactions between the image parts, ViT can recognize the intrinsic structures that exist in images [49].

ViT divides the input image into non-overlapping fixed-size patches and converts each patch into a feature representation, called patch embeddings. These patch embeddings are linearly transformed and serve as input tokens for the Transformer encoder [50]. Patches allow the model to observe the whole image capturing the global context. To encode positional information, positional embeddings are added to the patch embeddings [51]. This helps the model maintain awareness of the spatial arrangement of the patches in the original image. The Transformer encoder is the fundamental building block of the ViT architecture. It contains multiple layers of MSA and feed-forward neural networks. The MSA heads enable the model to capture both local and global relationships among patches, while the feed-forward networks introduce non-linearity to refine the features [52]. The output of the last encoder layer is typically pooled, and a classification head is added to make predictions. For image classification tasks, the pooled representation is often followed by a fully connected layer with softmax activation for class probabilities, whereas for image segmentation a specialized decoder block is used to obtain image masks [53].

# 4. ViTs for Medical Image Segmentation

CNNs are frequently utilized in medical image analysis for a variety of applications, such as tumor detection [54], [55], COVID-19 detection [56], skin lesion detection [57], and segmentation [58]. However, CNNs may struggle to learn explicit long-distance dependencies because of their limited receptive fields [59]. In contrast, ViT-based medical diagnostic systems can capture large receptive fields and have shown exemplary performance in various medical image-related tasks [60], [61]. A number of ViT-based systems have been developed for various medical image modalities, which include 1) classification-based systems [62], 2) detection-based systems [63], and 3) segmentation-based systems [64]–[66].

In medical imaging, the classification and detection of various types of cancerous cells is crucial to help pathologists carry out in-time diagnosis of disease. ViTs and more recently HVTs due to their compelling benefits over CNNs, have emerged as highly effective solutions [67]. ViT and HVT-based systems have shown remarkable performance across various medical image analysis tasks [68] including breast ultrasound image classification [69], COVID-19 detection [70]–[73], histopathology image analysis [74]–[78], mitosis detection [79]–[81], skin lesion detection [82]–[85]. These approaches have significantly improved the accuracy and efficiency of medical image segmentation and have the potential to enhance clinical diagnosis and decision-making [86], [87].

## 4.1. ViT-based Medical Image Segmentation Approaches

Several segmentation-based ViTs have been proposed which can be broadly classified based on their architectural modifications and the different training strategies they have employed. Most of the ViT-based medical image segmentation approaches utilize a UNet-like encoder-decoder architecture, where a ViT architecture is either employed in 1) the encoder, 2) in the decoder, or

3) in between encoder-decoder, or 4) both the encoder and decoder are ViT-based architectures. Details of each of these categories are discussed below.

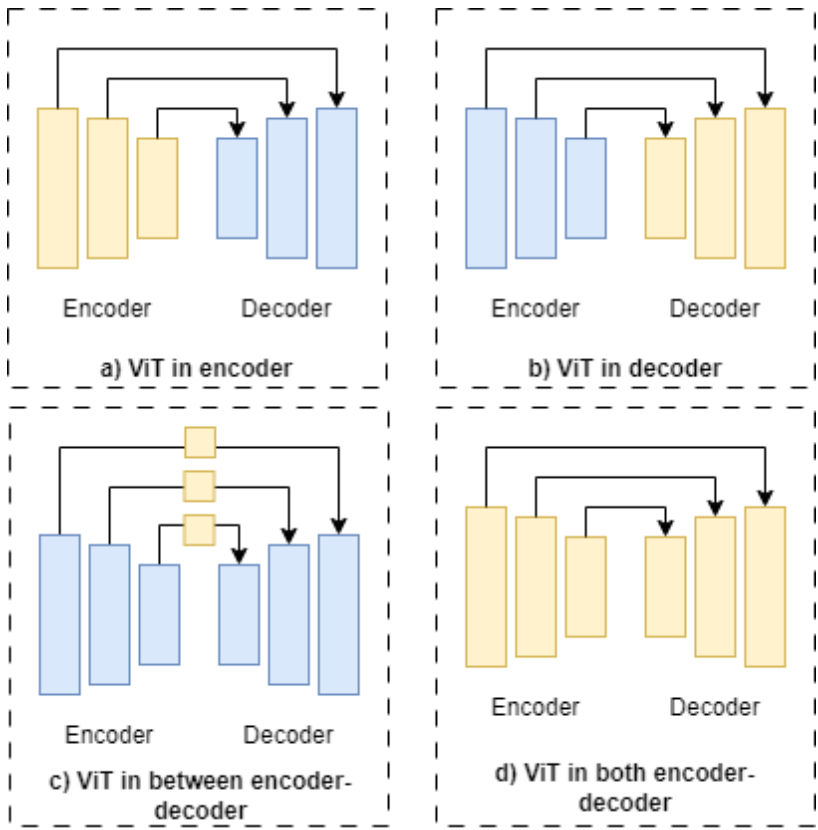

Figure 2: Taxonomy of ViT-based medical image segmentation approaches, where yellow colored layers are the ViT-based layers.

### 4.1.1. ViT in Encoder

An encoder-decoder architecture serves as the foundation for the majority of medical image segmentation techniques. The encoder architecture is responsible for learning the hidden embedding in the images, which are then decoded to a segmentation mask by a decoder. For an effective information transfer between the encoder and decoder, skip connections are also added along with a conventional sequential flow. These skip paths are either direct connections or CNN-based processing blocks. Many ViT-based medical image segmentation approaches utilize a ViT architecture in their encoder to incorporate global relationships in their learned latent space. Such techniques benefit from ViT's MSA to exploit global features while feature extraction in their encoder block. The CNN-based decoder then utilizes this high-level information for predicting pixel-level segmentation masks. In a study by Hatamizadeh et al, a Transformer-based architecture UNETR (UNet Transformer) is developed to perform 3D medical image segmentation by overcoming the locality-related limitations of UNet [88]. It has a ViT-based encoder to effectively capture multi-scale global information from input volumes. To compute the final semantic segmentation output, the encoder is linked directly to a decoder at different resolutions, via skip connections, similar to a U-Net.

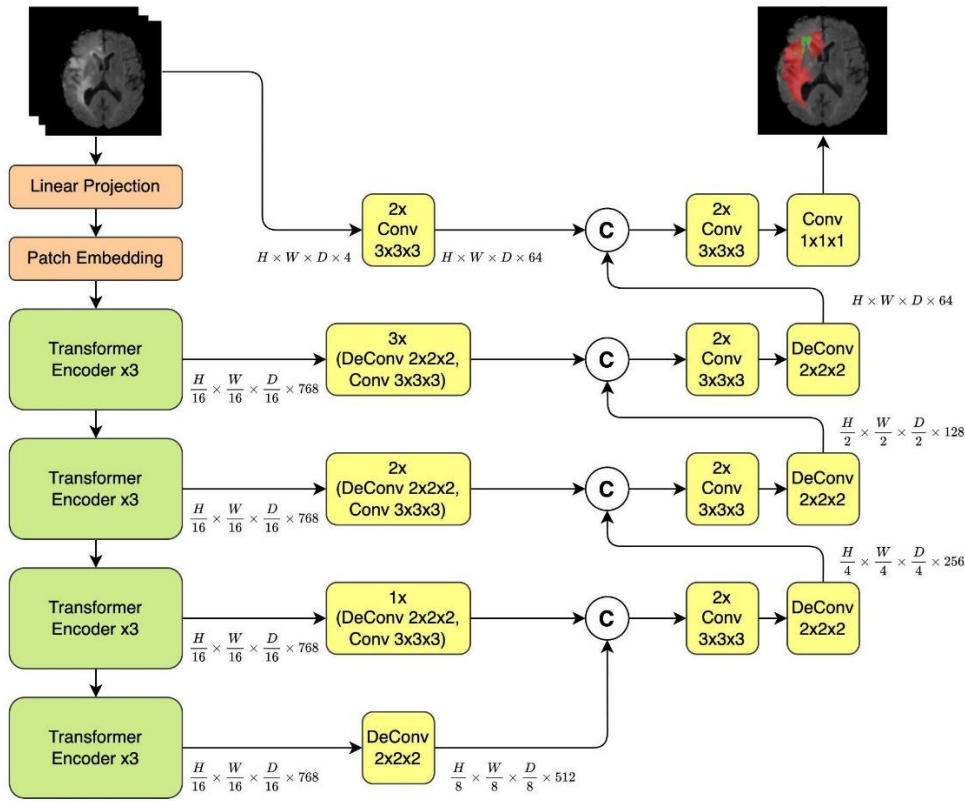

Figure 3: Architectural diagram of UNETR, where ViT is utilized in the encoder architecture.

### 4.1.2. ViT in Decoder

Some encoder-decoder techniques incorporate ViT in their decoder part only to predict accurate segmentation images [89]. As the global context is very important in predicting boundary-perfect segmentation masks and differentiating between background and object of interest, the MHSA is beneficial for such an objective during the decoding stage. ConvTransSeg [89] utilizes a CNN-based encoder for feature learning and a ViT-based decoder, interlinked at multi-stages. ConvTransSeg showed superior results on binary and multi-class segmentation problems, including skin lesions, polyps, cell, and brain tissue.

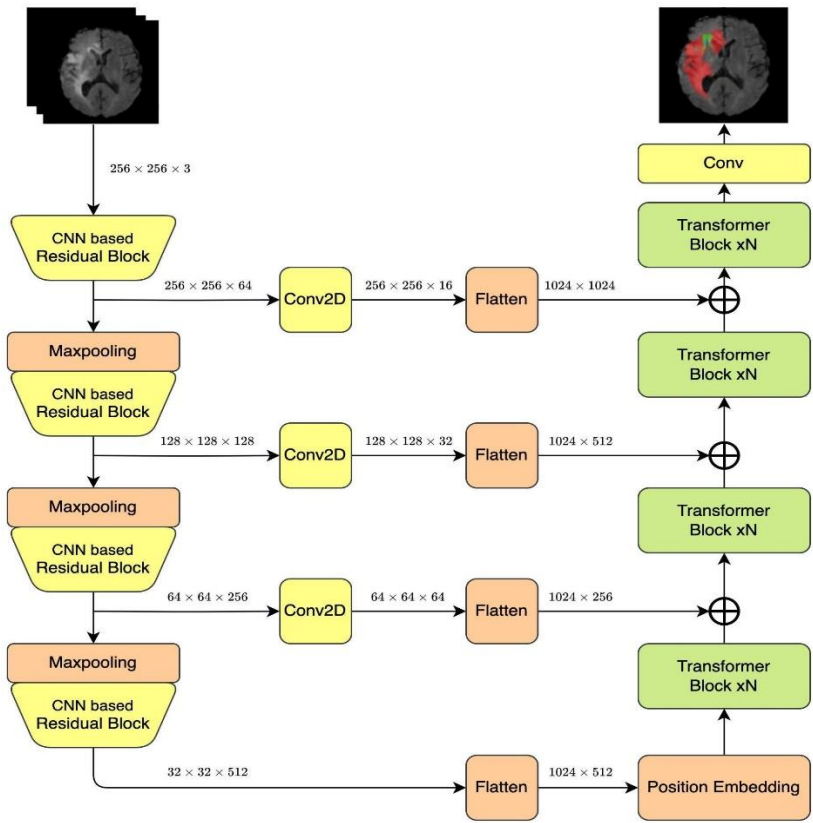

Figure 4: Architectural diagram of ConvTransSeg, where ViT is employed in Decoder.

### 4.1.3. ViT in both Encoder-Decoder

Various researchers have utilized ViTs in both the encoder and decoder architectures to fully exploit the attention mechanism. Coa et al. introduced Swin-Unet [90], to perform segmentation in medical images. Their architecture is based on Swin Transformer [91], with the encoder having a shifted-window-based approach and patch embedding layers in the decoder. Both encoder and decoder architectures are designed hierarchically to enhance segmentation accuracy and robust generalization. Zhou et al. proposed nnFormer [92], a volumetric Transformer network to carry out segmentation. The architecture utilizes attention-based skip connections to incorporate both local and long-term relationships in the volumetric images. Moreover, they utilized volume-based (V-MSA) and shifted version (SV-MSA) of multi-head self-attention to reduce the computational complexity and capture multi-scale information. MISSFormer [93] proposed by Huang et al., is an encoder-decoder architecture that incorporates the Enhanced Transformer Block and the Enhanced Transformer Context Bridge for the fusion of hierarchical features with less computational complexity. TransDeepLab [94] combines the DeepLabv3 network [95][90] with a shifted window-based Swin Transformer. It utilizes variable window sizes within the Swin-Transformer modules to fuse information at multiple scales.

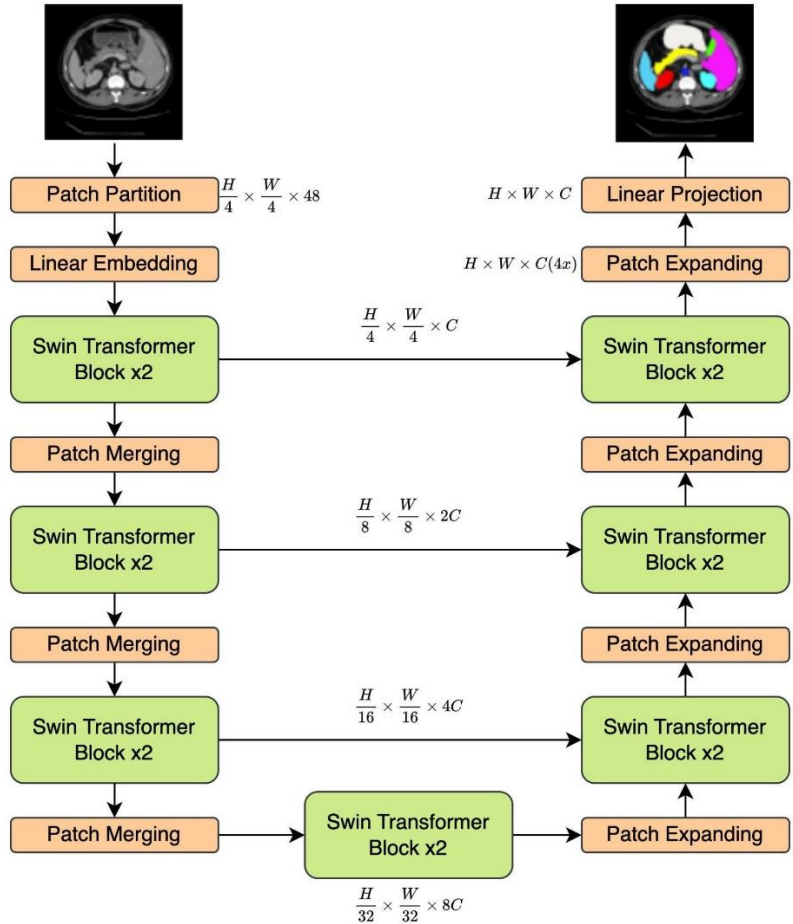

Figure 5: Swin-UNet architecture, with ViT in both Encoder-Decoder.

### 4.1.4. ViT in between Encoder-Decoder

Li et al. proposed ATTransUNet, an architecture based on ViTs utilizing adaptive tokens. Within their model, ViT is integrated into skip connections [96]. They utilized an Adaptive Token Extraction Module (ATEM) in their encoder to extract the most discriminating visual tokens in the images. This results in reduced complexity and enhanced performance.

In the decoder architecture, they employed a Selective Feature Reinforcement Module (SFRM) to focus on the most contributing features.

Dual Cross-Attention (DCA), a straightforward but powerful attention module that improves skip connections in U-Net-based architectures for medical image segmentation, was proposed by Ates and co-authors [97]. By modeling the channel and spatial relationships from the features obtained by the multi-scale encoder, DCA bridges the semantic gap between encoder and decoder features. DCA first employs the Channel Cross-Attention (CCA) and then the Spatial Cross-Attention (SCA) module to model channel-wise and spatial relationships. At last, the encoder features are upsampled to correspond with the corresponding layers of the decoder.

ViT-V-Net applies ViT blocks at the bottleneck layer between CNN-Encoder and CNN-Decoder for volumetric medical image registration task in an unsupervised manner [98]. CoTr provides an efficient hybrid architecture that uses CNN for feature extraction and a deformable self-attention mechanism for global context modeling [99]. The selective multi-scale deformable multi-head self-attention (MS-DMSA) is used, reducing computational complexity and enabling faster convergence.

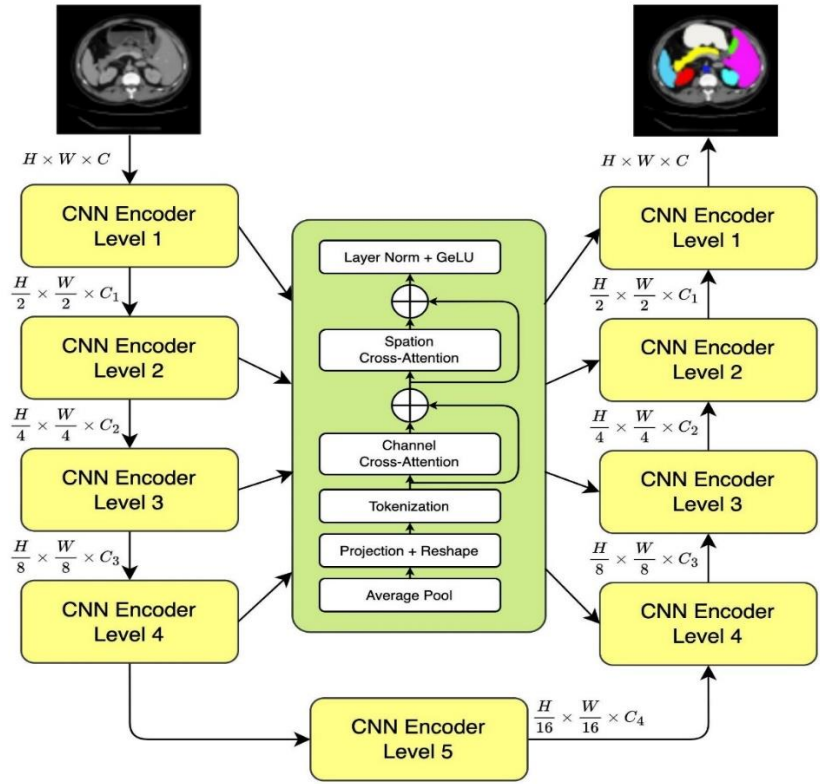

Figure 6: Architecture of the DCA Net, with ViT in between Encoder-Decoder

## 4.2. Hybrid ViT-Based Medical Image Segmentation Approaches

The pure ViT architectures, which rely exclusively on attention mechanisms and lack convolution operators, may miss low-level details, leading to inaccurate segmentation results. HVTs, by integrating the strengths of both ViTs and CNN architectures, demonstrate the ability to capture long-range as well as local context within the input data. This unique combination empowers HVTs to attain cutting-edge performance across various tasks, notably excelling in medical image segmentation [100], [101].

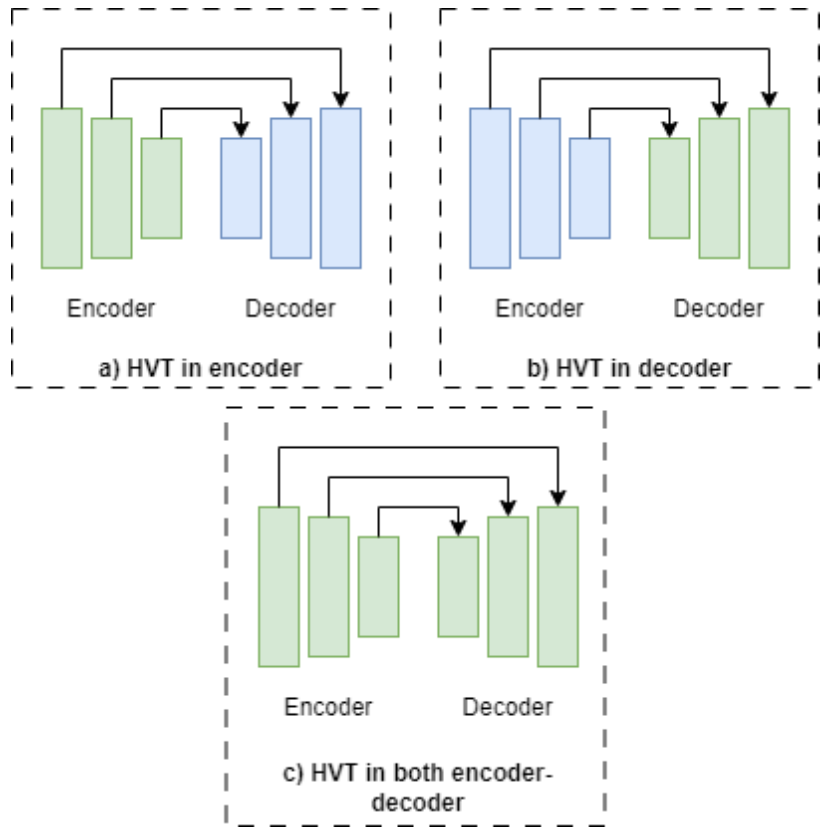

Figure 7: Taxonomy of various medical image segmentaton approaches based on HVTs, where green-color represents hybrid (CNN + ViT) layers.

### 4.2.1. Hybrid ViT in Encoder

The most common trend in recent research is to incorporate HVTs in the encoder stage. TransUNet utilizes an HVT architecture in its encoding stage and a cascaded upsampler in its decoding stage [102]. It combines the advantages of ViT and U-Net, capturing both local correlation and long-term relationships within the input images for accurate segmentation results. Wang et al. proposed a segmentation architecture named TransBTS that combines a ViT and a 3D CNN for global contextual information and local details, respectively [103]. It processes 3D medical volumetric data to capture local correlation and long-range dependencies within the image slices. TransFuse utilized a BiFusion module to combine its encoder and decoder branches which were CNN and ViT architectures, respectively [104]. MedT introduced a novel attention mechanism and training strategy for medical image segmentation [105]. It consists of a global module (ViT-based) and a local module (CNN-based) to capture high-level and fine details within the pixels. Swin-UNETR combines Swin Transformers with a U-shape architecture for the task of segmenting brain tumor regions [106]. This entails dividing the input into non-overlapping patches and utilizing windowing mechanisms. However, the integration of the self-attention mechanism into CNNs may result in high computational complexity due to the large spatial size. In this regard, H2Former is introduced to efficiently combine the strengths of MSA and CNNs to perform medical image segmentation [107]. H2Former outperformed the previous techniques by maintaining its computational efficiency in terms of model parameters, Floating Point Operations (FLOPs), and inference time.

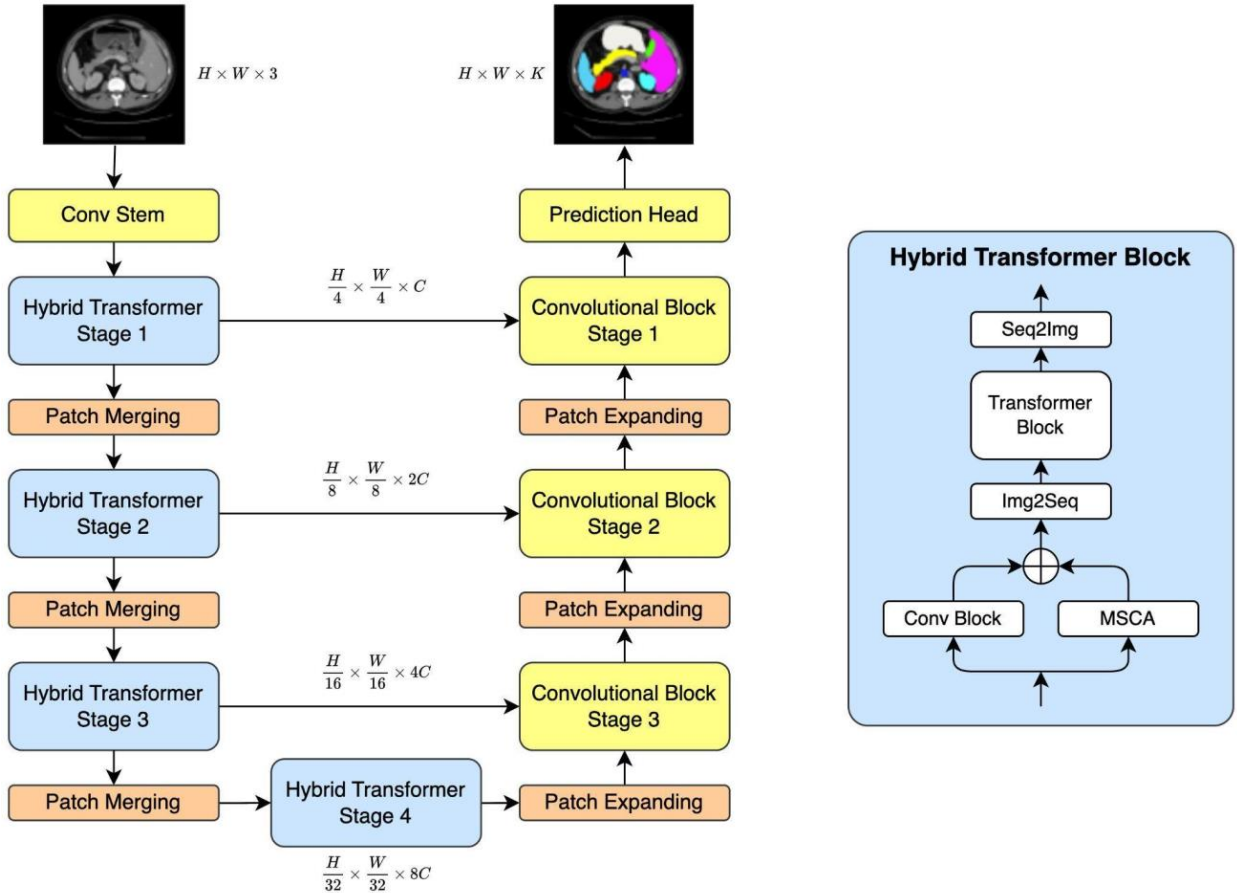

Figure 8: Architectural diagram of TransUNet, where a hybrid ViT is utilized in the encoder.

### 4.2.2. Hybrid ViT in Decoder

To leverage the benefits of Hybrid ViT in decoding stages, recent research incorporates an HVT in the decoder architecture. A recent study presents a unified framework, UNetFormer, employing a 3D Swin Transformer on the encoding side and a combination of CNN and Transformer on the decoding side [108].

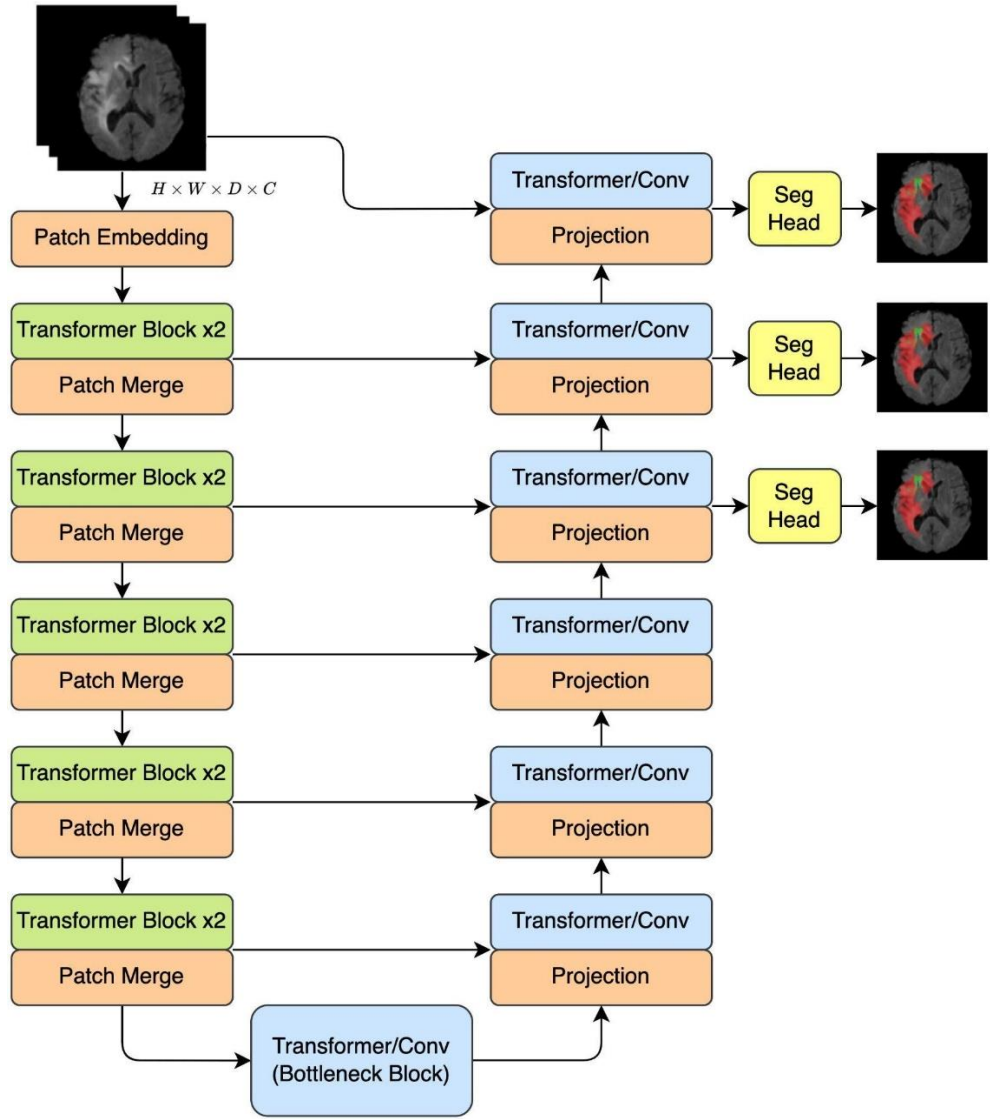

Figure 9: Architecture of UNetFormer, where in the decoder architecture a hybrid ViT is employed.

### 4.2.3. Hybrid ViT in both Encoder-Decoder

Many researchers utilize ViT and CNN-based modules in both the encoder and decoder parts of their architecture [109]. Recently, MaxViT-UNet has been developed as an innovative HVT-based decoder specifically for medical image segmentation [110]. The authors effectively utilized the multi-axis self-attention mechanism which allows the model to attend to features along local and global axes, enhancing the discriminative capacity between object and background regions, thus improving segmentation efficiency [111].

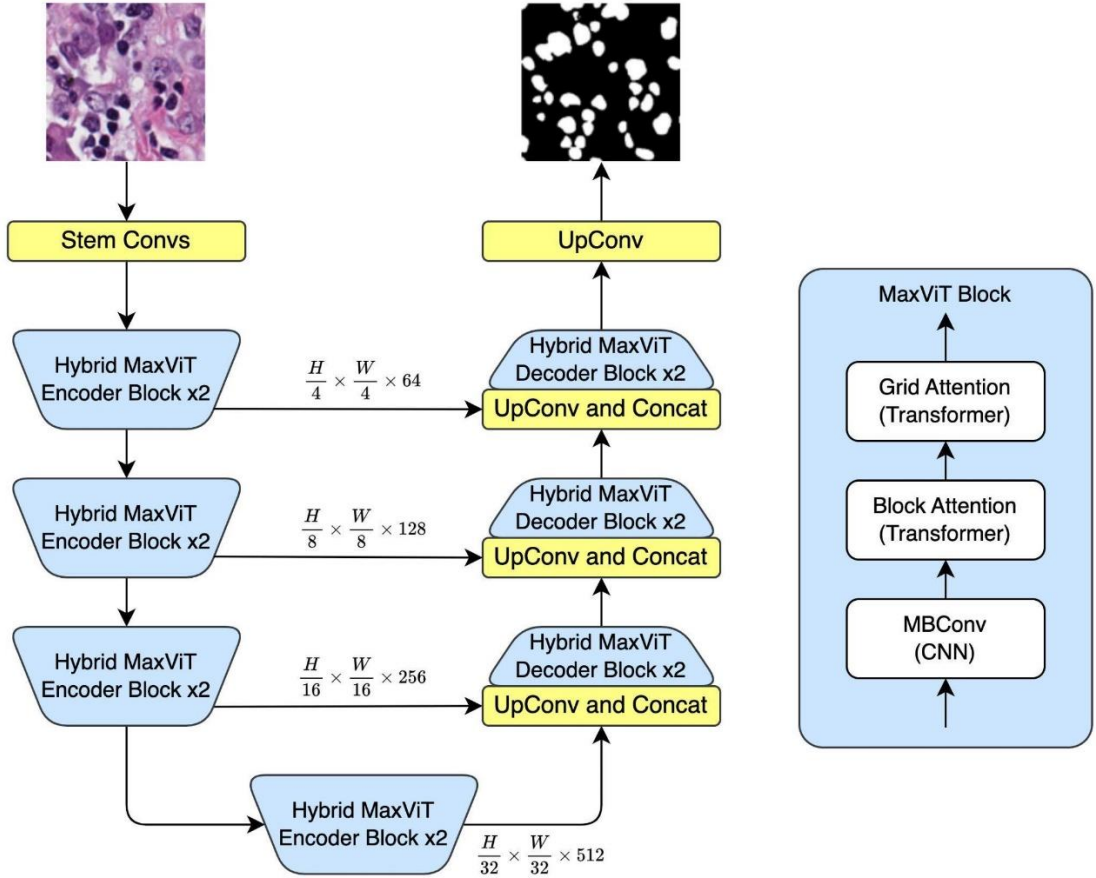

Figure 10: Architecture of MaxViT-UNet, with hybrid ViT in both encoder and decoder.

## 5. ViT-based Medical Image Segmentation Applications

Due to their ability to capture global relationships between pixels, ViTs have sparked exponential growth in medical image analysis. ViT-based medical image segmentation can be broadly categorized based on the specific image modality. These methods include but are not limited to, medical image segmentation in 1) CT images, 2) Histopathological images, 3) Microscopy images, 4) MRI images, 5) Ultrasound, and 6) X-Ray images. Medical image processing is the primary step in medical analysis as it facilitates diagnostic proficiency through various tasks including cell counting, classification, detection, and segmentation. However, medical image segmentation is the most commonly used task in medical diagnosis [112]–[114].

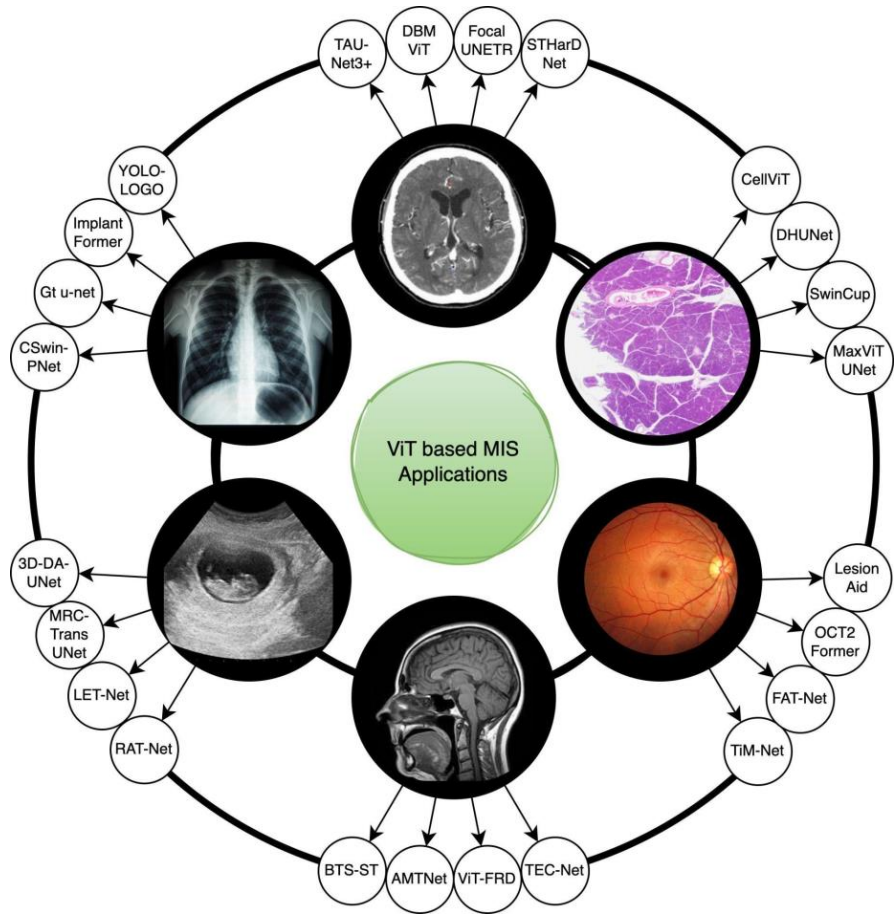

Figure 11: Taxonomy for Section 5 highlighting recent ViT-based models for different modalities

## 5.1.   CT Images

CT (Computed Tomography) is an advanced medical imaging technique to produce detailed cross-sectional body images using X-rays for disease detection and diagnosis. Currently, most segmentation techniques for CT images make use of CNNs, but recent progress of ViTs has shifted the trend, and many frameworks have emerged for CT modality [115]–[118].

TAU-Net3+ replaced UNet's CNN encoder with a ViT for kidney tumor segmentation, and a novel attention mechanism called an encoder-decoder Transformer (EDformer) was added to the skip connection to learn local characteristics [119].

In response to the challenges encountered by conventional methodologies in achieving accurate prostate segmentation in CT images, the FocalUNETR has introduced an innovative image segmentation framework leveraging a focal Transformer [117]. This architectural design proficiently captures both localized visual features and comprehensive contextual information with efficiency and effectiveness. The authors have innovatively incorporated an auxiliary module known as "boundary-induced label regression," which serves to complement the principal task of prostate segmentation. Hoa and co-authors proposed DBM-ViT, for the discrimination of COVID-19, and other pneumonia types by utilizing CT and Chest X-ray images [120]. The utilization of depthwise convolutions with diverse expansion rates within DBM-ViT enhances the adept capture of global information. STHarDNet was developed by Gu and co-authors to perform cerebral hemorrhage segmentation [121]. In their approach, HarDNet and Swin Transformer were integrated to enhance the segmentation performance beyond what each model can achieve independently.

*Table 1: Different ViT and HViT-based techniques for CT Image Modality*

| Method | Architecture | Highlights | Performance |
|---|---|---|---|
| TA-UNet3+ (2023) [119] | ViT (in Encoder and in between Encoder-Decoder) | By integrating Transformer and U-Net, a new network architecture is suggested. The new attention mechanism, EDformer, is also added to the skip connection. | KiTS19 Dataset Dice: 0.9638 (tumor) Dice: 0.9885 (kidney) |
| STHarDNet (2023) [121] | ViT (in between Encoder and Decoder) | Combined the HarDNet and Swin Transformer to propose a hybrid model for cerebral hemorrhage segmentation using 2D CT images. | AIHub CT Dataset Dice: 0.7280 IoU: 0.6120 |
| FocalUNETR (2023) [117] (code) | ViT (Encoder-based) | It adeptly captures both local and global features and incorporates an auxiliary label regression task triggered by boundaries to handle indistinct edges in CT images. | AMOS Dataset Dice: 0.8379 Private Dataset Dice: 0.8923 |

## 5.2. Histopathological Images

Histopathological images, obtained from the biopsy tissue samples, are analyzed by pathologists to investigate cellular architecture and spot microscopic abnormalities or illnesses [122]. Histopathology-based analysis is crucial for timely cancer diagnosis and prevention. While the use of ViTs on histopathology images is still in its early stages, nevertheless, they have already given promising results on various histopathology segmentation tasks [123]–[125].

Wang et al. introduced DHUnet, an innovative feature fusion approach that blends global and local features for the segmentation of WSI. This method incorporates Swin Transformer and ConvNeXt modules within a dual-branch hierarchical U-shaped architecture [126].

In another recent approach, the Swin Transformer with Cascaded UPsampling (SwinCup) was introduced to segment histopathology images [127]. To improve the feature aggregation of the proposed SwinCup, the authors introduced a cascaded up-sampling decoder in conjunction with the encoder.

In another technique, MaxViT-UNet, the authors proposed a Unet-like encoder-decoder CNN-Transformer framework [110]. The proposed hybrid decoder utilized Multi-Axis Self-Attention (Max-SA) to improve the segmentation results. These models subsequently have the potential to contribute to patient stratification in clinical trials, patient selection for targeted therapy, and personalized treatment planning.

*Table 1: Different ViT and HViT-based techniques for Histopathological Image Modality*

| Method | Architecture | Highlights | Performance |
|---|---|---|---|
| CellViT (2023) [37] (code) | ViT (Encoder-Based) | Devised a deep learning architecture centered around Vision Transformer to enable automated instance segmentation of cell nuclei in digitized tissue samples. | PanNuke F1-score: 0.8300 |
| Swin-MIL (2022) [115] | ViT (Encoder-Based) | Introduced Transformer into the Multi-Instance Learning framework (MIL) to capture global | Colon Cancer F1-score: 0.8500 |

| (code) | | dependencies. | |
|---|---|---|---|
| DHUnet (2023) [126] (code) | Hybrid ViT (Encoder-Based) | Employed a Unet-based architecture with a Swin Transformer-based branch and a ConvNeXt branch to integrate both local and global features to segment Whole Slide Images (WSI). | Liver Dataset<br>Dice: 93.07<br>Jaccard: 87.04<br><br>WSSS4LUAD<br>Dice: 86.54<br>Jaccard: 77.87<br><br>BCSS<br>Dice: 78.88<br>Jaccard: 68.01 |
| MaxViT-UNet (2023) [110] (code) | Hybrid-ViT (in both Encoder and Decoder) | The MaxViT-UNet leverages the hybrid MaxViT block in all encoder and decoder stages. The MaxViT block, consisting of MBConv, Block Attention and Grid Attention, effectively extracts local and global features. | MoNuSeg18<br>Dice: 0.8378<br>IoU: 0.7208<br><br>MoNuSAC20<br>Dice: 0.8215<br>IoU: 0.7030 |

## 5.3.    Microscopy Images

Microscopy involves using microscopes to view small objects, such as cells, tissues, and microorganisms [128]. Different types of microscopy, such as light microscopy, electron microscopy, and fluorescence microscopy, offer various levels of detail and resolution [129], [130]. The segmentation of skin lesions is a critical step in computer-aided diagnosis and therapy planning. It enables the objective identification and measurement of lesion size, shape, and characteristics, playing a vital role in classifying benign and malignant lesions, early detection, and monitoring temporal changes. Drawing inspiration from the successful application of ViTs in various medical domains, recent methodologies have proposed ViT-based solutions for the segmentation of skin lesions [82]–[84]. Notably, in the LesionAid framework [84], the authors introduced a novel multi-class prediction approach for skin lesions, employing ViT and ViTGAN [131]. To mitigate issues related to class imbalance, ViT-based Generative Adversarial Networks (GANs) are employed in these techniques.

Alterations in the structure and function of retinal blood vessels have been associated with cardiovascular conditions, including coronary heart disease, atherosclerosis, and hypertension. The automated segmentation of retinal images and subsequent analysis are essential for the evaluation and prediction of related diseases, thereby contributing to public health. Various research approaches have creatively harnessed the capabilities of Transformers in the context of retinal vessel segmentation in prior studies [132]–[138]. The OCT2Former [135], characterized by an encoder-decoder architecture, employs a dynamic Transformer encoder in conjunction with a lightweight decoder. The dynamic token aggregation Transformer within the dynamic transformer encoder is instrumental in capturing global context information about retinal vessels, while an auxiliary convolution branch addresses the Transformer's inherent lack of inductive bias. Du et al. introduced an ensemble strategy that integrates different foundational deep learning models, such as the FCN-Transformer and Pyramid Vision Transformer (PVT), for retinal vessel segmentation [134].

*Table 2: Different ViT and HViT-based techniques for Microscopy Image Modality*

| Method | Architecture | Highlights | Performance |
|---|---|---|---|
| FAT-net (2022) [82] (code) | Hybrid ViT (Encoder-Based) | Implemented a dual encoder comprising both CNNs and Transformer branches to simultaneously gather local characteristics and global context information. | ISIC 2016 Dice: 0.9159 ISIC 2017 Dice: 0.8500 ISIC 2018 Dice: 0.8903 PH2 Dice: 0.9440 |
| Fully Transformer Network (2022) [83] (code) | ViT (in both Encoder-Decoder) | Offered a self-attention paradigm for image segmentation and other downstream tasks. | ISIC 2018 Dice: 0.9000 Jaccard: 0.8290 |
| CAViT (2022) [127] | ViT (in between | Introduced channel attention vision Transformer that jointly applies the | DRIVE Accuracy: 0.9700% |

| | Encoder-Decoder) | efficient channel attention and the vision Transformer. | CHASE DB1<br>Accuracy: 0.9761% |
|---|---|---|---|
| TiM-Net (2022)<br>[138]<br>(code) | ViT<br>(in between Encoder-Decoder) | Integrated M-Net, various attention mechanisms, and weighted side output layers to create TiM-Net for the segmentation of retinal vessels. | STARE<br>Accuracy: 0.9711%<br>AUC: 0.9670<br><br>CHASE DB1<br>Accuracy: 0.9711%<br>AUC: 0.9648<br><br>DRIVE<br>Accuracy: 0.9638%<br>AUC: 0.9682 |
| PCAT-UNET (2022)<br>[136] | Hybrid ViT<br>(in both Encoder-Decoder) | Proposed an encoder-decoder network leveraging both ViT and Convolution and incorporating skip connections to merge deep and shallow features from both modules. | DRIVE Dataset<br>F1-score: 0.8160<br>Accuracy: 0.9622%<br>AUC: 0.9872<br><br>CHASE DB1<br>F1-score: 0.8273<br>Accuracy: 0.9812%<br>AUC: 0.9925<br><br>STARE Dataset<br>F1-score: 0.8836<br>Accuracy: 0.9796%<br>AUC: 0.9953 |
| OCT2Former (2023)<br>[135] | Hybrid ViT<br>(Encoder-based) | Introduced OCT2Former, employing an encoder-decoder structure that comprises a dynamic Transformer encoder and a lightweight decoder primarily. | OCTA-SS<br>Jaccard: 0.8344<br><br>ROSE-1<br>Jaccard: 0.7855<br><br>OCTA-6M<br>Jaccard: 0.8099<br><br>OCTA-3M<br>Jaccard: 0.8513 |
| Axial Transformer and Convolutional Neural Network (2022)<br>[112] | Hybrid ViT<br>(Encoder-based) | Both the Axial Transformer and CNN are used to extract the global local features of the image, respectively. | CHASE DB1<br>F1-score: 0.8095<br>Accuracy: 0.9657% |
| MTPA_Unet (2022)<br>[132] | Hybrid ViT<br>(Encoder-based) | The Transformer is fused with the CNN, and the TPA module is used to learn both local and global-level information. | DRIVE<br>Accuracy: 0.9718%<br>Dice: 0.8318 |

| | | | CHASE DB1<br>Accuracy: 0.9762%<br>Dice: 0.8164<br><br>STARE<br>Accuracy: 0.9773%<br>Dice: 0.8557 |
|---|---|---|---|
| Group Transformer Network (GT U-Net) (2021) [139] [46] (code) | Hybrid ViT (in both Encoder-Decoder) | Utilized both the convolution and Transformer networks without pre-training weights. | DRIVE F1-score: 0.8458 |

## 5.4. MRI Images

Magnetic Resonance Imaging (MRI) generates intricate visual representations of tissues and organs within the human body through the application of robust magnetic fields. This imaging modality is particularly valuable in the examination of anatomical regions such as the joints, muscles, and vital organs like the heart and liver, spinal cord, and brain.

In recent times, there has been a notable increase in the utilization of ViT-based methodologies for the segmentation of cardiac magnetic resonance imaging (MRI) data, as evidenced by approaches such as [140]–[145]. Fan et al. introduced a knowledge refinement technique called Visual Transformer with Feature Recombination and Feature Distillation (ViT-FRD)[143]. In their approach, ViT functions as the student network, assimilating knowledge from a CNN and acting as the teacher network through optimized distillation losses. ViT-FRD incorporates two enhancements to improve training efficiency and efficacy.

The advancement of an automated and accurate methodology for brain tumor segmentation has the potential to expedite diagnosis across a range of oncologic phenotypes. In recent times, there has been a notable increase in the introduction of CNN layers in the Transformer-based models for the efficient segmentation of brain tumors, as evidenced by several studies [35]–[37], [103], [106], [108], [146]–[148]. One such model, DenseTrans [149], creatively amalgamates the Swin Transformer with an enhanced UNet++ network. This integration aims to extract local features from the convolutional layer and global features from the shift window operation and self-attention mechanisms of the Swin Transformer within the high-resolution layer. In the case of 3DCATBraTs [148], researchers adapted the Swin Transformer for the segmentation of brain tumors in 3D MRI images. This Transformer incorporates a modified CNN-encoder architecture, characterized by residual blocks and a channel attention module. Furthermore, in a different investigation, a robust 3D fusion segmentation network named AMTNet [150] was introduced, building upon the conventional U-shaped structure. Notably, AMTNet incorporates a Transformer-based feature fusion module designed to enhance the integration of multimodality features.

In the realm of breast lesion segmentation utilizing MRI images, as evidenced by several recent studies [151]–[153], Iqbal et al. introduced BTS-ST [152], which is inspired by the Swin Transformer, to enhance the feature representation capability, particularly for irregularly shaped tumors. BTS-ST innovatively incorporated pixel-level correlations within the Swin Transformer block to encode spatial knowledge. To address information loss, a Feature Compression Block was introduced, and a Relationship Aggregation block enabled the cohesive hierarchical combination of global dependencies from the Swin Transformer with features from CNN. In a separate study, Muller Franzes et al. introduced TraBS, enhancing the original SwinUNETR model for breast segmentation in multi-institutional MRI data [106]. TraBS strategically utilizes non-isotropic kernels and strides in the initial two stages to maintain a consistent depth, and it integrates deep supervision for the lower-resolution layers.

*Table 3: Different ViT and HViT-based techniques for MRI Image Modality*

| Method | Architecture | Highlights | Performance |
|---|---|---|---|
| TransBTS (2021) | ViT | Combined a 3D CNN to capture local spatial | BraTS 2019 |
| [103]<br>([code](#)) | (in between Encoder-Decoder) | features with Transformers to acquire global context and utilizes progressive upsampling in the decoder. | Dice: 0.7893 (ET)<br>Dice: 0.9000 (WT)<br>Dice: 0.8194 (TC)<br><br>BraTS2020<br>Dice: 0.7873 (ET)<br>Dice: 0.9009 (WT)<br>Dice: 0.8173 (TC) |
| BiTr-UNet (2021) [35]<br>([code](#)) | ViT<br>(in between Encoder-Decoder) | Used two Transformer layers and an attention component to improve features from the encoding and decoding units. | BraTS2021<br>Dice: 0.8874 (ET)<br>Dice: 0.9257 (WT)<br>Dice: 0.9350 (TC) |
| VT-UNet (2022) [36]<br>([code](#)) | ViT<br>(in both Encoder-Decoder) | Utilized two ViT layers in the encoder along with the window-based and cross-attention modules to learn global and local relationships. | BraTS2021<br>Dice: 0.8220 (ET)<br>Dice: 0.9190 (WT)<br>Dice: 0.8720 (TC) |
| Swin UNETR (2021) [106]<br>([code](#)) | ViT<br>(Encoder-Based) | Utilized Swin Transformer in the encoding unit and a CNN in the decoder. | BraTS2021<br>Dice: 0.8530 (ET)<br>Dice: 0.9270 (WT)<br>Dice: 0.8760 (TC) |
| SwinBTS (2022) [154]<br>([code](#)) | Hybrid ViT<br>(in both Encoder-Decoder) | SwinBTS utilized a 3D Swin Transformer for local and global-level feature learning along with the convolutional operations for up and down sampling. | BraTS2019<br>Dice: 0.7443 (ET)<br>Dice: 0.8975 (WT)<br>Dice: 0.7928 (TC)<br><br>BraTS2020<br>Dice: 0.7736 (ET)<br>Dice: 0.8906(WT)<br>Dice: 0.8030 (TC)<br><br>BraTS2021<br>Dice: 0.8321 (ET)<br>Dice: 0.9183 (WT)<br>Dice: 0.8475 (TC) |
| UnetFormer (2022) [108]<br>([code](#)) | Hybrid ViT<br>(Decoder-Based) | Combined 3D Swin Transformer-based encoder and CNN and Transformer-based decoders, and links the encoder with the decoder via skip connections at five different resolutions. | BraTS2021<br>Dice: 88.80 (ET)<br>Dice: 93.22 (WT)<br>Dice: 92.10 (TC) |
| CATBraTs (2023) [148] | Hybrid ViT<br>(Encoder-Based) | Proposed 3D Transformer based on Swin Transformer with a modified CNN-encoder architecture. | BraTS2021<br>Dice: 0.8260 (ET)<br>Dice: 0.8760 (WT)<br>Dice: 0.7990 (TC) |
| AMTNet (2023) [150] | Hybrid ViT<br>(in both Encoder-Decoder) | Introduced Transformer-based feature fusion module designed to better fuse multimodality features. | BraTS2021<br>Dice: 0.7340 (ET)<br>Dice: 0.9240 (WT)<br>Dice: 0.8950 (TC) |
| DenseTrans (2023) | ViT | Combined Swin Transformer with the | BraTS2021 |

| [149] | (in between Encoder-Decoder) | improved UNet++ network. | Dice: 0.8830 (ET) Dice: 0.9320 (WT) Dice: 0.8620 (TC) |
|---|---|---|---|
| TransConver (2022) [147] | Hybrid ViT (Encoder Based, in between Encoder-Decoder) | Proposed a parallel module, which uses convolution blocks and Transformer blocks, respectively, to extract local and global information. | BraTS2018 Dice: 0.8173 (ET) Dice: 0.9157 (WT) Dice: 0.8668 (TC) BraTS2019 Dice: 0.7840 (ET) Dice: 0.9019 (WT) Dice: 0.8257 (TC) |
| Focal Cross-Transformer (2023) [146] | ViT (Encoder Based) | Parallelized self-attention of horizontal and vertical fringes in the cross window to enlarge the receptive field. | BraTS2021 Dice: 0.8728 (ET) Dice: 0.9328 (WT) Dice: 0.8735 (TC) |
| Segtran (2021) [155] (code) | ViT (in between Encoder-Decoder) | Incorporated squeezed attention block to regularize self-attention of Transformers and expansion block to learn diversified representations. | BraTS2019 Dice: 0.7400 (ET) Dice: 0.8950 (WT) Dice: 0.8170 (TC) |
| BTS-ST (2023) [152] | Hybrid ViT (Encoder Based) | Incorporated Swin-Transformer into traditional CNNs-based U-Net. Also, encoded spatial knowledge in the Swin Transformer block by developing pixel-level correlation to improve the feature representation capability of irregularly shaped tumors. | MRI Dataset F1-Score: 0.8330 |
| TraBS (2023) [153] | ViT (Encoder Based) | Modified SwinUNETR and added deep supervision for the lower-resolution layers. | UKA Dice: 0.9160 DUKE Dice: 0.8640 |

## 5.5. Ultrasound Images

Ultrasound or sonography, employs high-frequency sound waves to produce real-time images of organs and tissues. Extensively used in obstetrics to monitor pregnancies and evaluate fetal development, ultrasound holds a pivotal position in healthcare. The early detection of breast cancer has shown the potential to reduce fatality rates by over 40%, highlighting the substantial importance of automated breast tumor detection for healthcare practitioners.

Zhu et al. [156] introduced the region-aware Transformer network (RAT-Net), which skillfully integrates information from breast tumor regions at various scales, leading to accurate segmentation. Likewise, Liu et al. [157] devised a hybrid architecture that incorporates Transformer layers into the decoder segment of a 3D UNet, facilitating efficient tumor segmentation in volumetric breast data. The LET-Net architecture [158], introduced recently, combines Transformer and convolutional approaches. Its feature-aligned local enhancement module facilitates the extraction of distinctive local features while maintaining alignment with adjacent-level features. Additionally, a progressive local-induced decoder focuses on recovering high-resolution spatial details through a series of local reconstruction and refinement stages, guided by adaptive reconstruction kernels and enhanced through a split-attention mechanism.

Similarly, MRC-TransUNet [159] introduced an innovative approach by merging Transformer and UNet components. It incorporates a lightweight MR-ViT to bridge the semantic gap and utilizes a reciprocal attention module (RPA) to counteract potential detail loss. These research endeavors collectively suggest that techniques based on ViTs have the potential to significantly enhance the accuracy of medical image segmentation, holding promise for various clinical applications.

*Table 4: Different ViT and HViT-based techniques for Ultrasound Image Modality*

| Method | Architecture | Highlights | Performance |
|--------|-------------|-----------|-------------|
| RAT-Net (2022) [156] (code) | Hybrid ViT (Encoder Based, in between Encoder-Decoder) | Captured multi-scale information for breast cancer segmentation by utilizing a region-aware Transformer network. | ABUS Accuracy: 0.8451% Dice: 0.6833 IoU: 0.5865 |
| 3D-UNet (2021) [157] | Hybrid ViT (in between Encoder-Decoder) | Adopted a structure based on 3D U-Net with attention mechanism and Transformer layers to optimize the extracted image features. | ABVS Dice: 0.7636 IoU: 0.6214 |
| Cswin-Pnet (2023) [160] | Hybrid ViT (Encoder Based, in between Encoder-Decoder) | Connected CNN and Swin Transformer as a feature extraction backbone to construct a pyramid-structured network for effective feature encoding and decoding. | Dataset 1 Dice: 0.8725 Dataset 2 Dice: 0.8368 |
| BTS-ST (2023) | Hybrid ViT | Integrated the Swin Transformer into a | Ultrasound Dataset |

| [152] | (Encoder Based) | conventional CNN-based U-Net to improve the feature learning ability for irregularly shaped tumors. | F1-Score: 0.9080 |

## 5.6. X-Ray Images

X-ray images serve as a cornerstone for both diagnostic and therapeutic reasons and offer priceless insights into the interior organs of the human body. Modern deep learning models can be effectively used to diagnose a variety of medical diseases using X-ray pictures, which are well known for their important role in this process.

Tooth root segmentation is an important stage in dental image analysis as it allows dentists to precisely measure the size and shape of tooth roots and discover any abnormalities that may be present. Many recent works have employed ViTs for tooth root segmentation [139], [161], [162]. Yang et al. introduced ImplantFormer, an implant position regression network built on Transformers that uses oral CBCT data to predict implant location automatically [161]. ImplantFormer uses a 2D axial picture of the tooth crown area to forecast the implant position before fitting a centerline to precisely locate the implant at the tooth root. In another study [163], Sheng et al. introduced SwinUnet, a U-shaped Transformer-based architecture with an encoder, decoder, and skip-connections, designed specifically for panoramic radiograph segmentation.

Mammography is a specific type of X-ray used to examine breast tissue for signs of breast cancer. Certain research studies have directed their efforts toward refining architectural designs for improved segmentation of irregular tumor boundaries [160], [164]. CSwin-PNet was recently proposed for the task of breast lesion segmentation. Its architecture is based on a Pyramid Network that integrates the CNN and Swin Transformer [160].

*Table 5: Different ViT and HViT-based techniques for Ultrasound Image Modality*

| Method | Architecture | Highlights | Performance |
| --- | --- | --- | --- |
| GT U-Net (2021) [139] (code) | Hybrid ViT (in both Encoder-Decoder) | Utilized both the CNN and Transformer architectures without pre-training weights. | ToothRoot Dice: 0.9254 |
| ImplantFormer (2022) [161] | ViT (Encoder-Based) | The prediction network is trained using the 2D axial view of the tooth crown, which is easily visible and well-captured through CT imaging as it is exposed in the air | CBCT Dataset AP@0.75: 0.1370 |
| SWin-UNet (2023) [163] | ViT (in both Encoder-Decoder) | Implemented SWin-Unet CNN for tooth segmentation on panoramic radiographs. | PLAGH-BH F1-score: 0.6372 PDX F1-score: 0.8203 |
| YOLO-LOGO (2022) [164] | ViT (Encoder-Based) | The process involves YoloV5L6 detecting and isolating breast masses in mammograms. A modified LOGO is applied separately to the entire and cropped images on global and local Transformer branches. The final segmentation decision is formed by merging the outcomes from both branches. | CBIS-DDSM mAP: 0.6500 INBreast mAP: 0.6140 |

## 6. Challenges

ViTs have shown great potential for revolutionizing medical imaging, offering an innovative approach to analyzing and extracting valuable data from medical images. Despite their promise, ViTs encounter several challenges in accurate medical image segmentation, which include:

- Unlike CNNs, ViTs rely heavily on self-attention processes, neglecting explicit incorporation of local features or receptive fields. Various techniques, including HVTs, aim to overcome this limitation by capturing both global and local information simultaneously.

- In addition, when dealing with large and detailed 3D medical images using ViTs, there are difficulties related to scalability and computational demands. This requires making modifications to the model size, decoding sequence length, and employing parallelization methods.

- Medical datasets may pose unique challenges due to data scarcity, imbalanced classes, and domain shift concerns. Researchers address these issues through synthetic data generation, data augmentation, data balancing, and novel training schemes to improve the model generalization. However, ensuring the robustness of models during deployment on previously unseen cases remains a challenge in the medical domain.

- The quality of annotations in medical image analysis is crucial for training reliable models, but it is hindered by time-consuming and costly processes conducted by radiologists and pathologists. Even with expert annotations, variations and inconsistencies can occur, negatively impacting the model's learning process and performance.

Addressing these challenges, along with the lack of interpretability in ViTs, remains essential for their effective application in healthcare. Improving interpretability, ensuring generalizability across diverse datasets, and implementing solutions for privacy concerns and adversarial attacks are vital steps in promoting the accurate and reliable use of ViTs in medical image segmentation for automated diagnosis.

## 7. Future Recommendations

Dealing with the specific challenges faced by ViTs in medical image segmentation presents several avenues for future research and innovation.

- One key direction involves the development of new hybrid architectures by combining the strengths of ViTs and CNNs. This hybrid approach aims to strike a balance between capturing global structural relationships and local details crucial for precise medical imaging segmentations, facilitating the deployment of real-world automated diagnostic systems.

- Improving training paradigms and methodology for ViTs in medical applications is another focus area. Fine-tuning on public datasets, incorporating relevant constraints, and employing adaptive methods to address biases and variations in class distributions are crucial steps. Additionally, complex reinforcement learning algorithms and tailored data augmentation techniques can enhance ViTs' capabilities to handle diverse pathological properties within medical images.

- Exploring ensemble methods emerges as a fascinating approach, involving the combination of predictions from multiple diverse ViT backbones. By leveraging various models with different specifications and architectures, this collaborative approach aims to enhance resilience, versatility, and overall efficiency in medical visualization tasks, reducing reliance on individual networks.

- Incorporating domain knowledge is emphasized to make deep learning more effective, especially in medical imaging. Techniques such as attention mechanisms can be fine-tuned to focus on critical areas, and the integration of anatomical priors or handcrafted features can guide the model. This incorporation of domain knowledge from medical professionals enhances segmentation accuracy and reliability in medical image analysis.

- Transfer learning stands out as a promising approach to address data scarcity challenges. Pre-training ViTs on large-scale natural image datasets and fine-tuning them on specific medical imaging tasks allows the models to leverage knowledge from diverse images. This approach enables ViTs to generalize better, improve performance, and potentially reduce the need for extensive labelled medical datasets.

- Lastly, multi-modal learning is identified as an exciting opportunity for ViTs in medical image segmentation. ViTs, with their ability to handle sequence data, can effectively accommodate inputs from various modalities, capturing dependencies and interactions between different data sources. This multi-modal capability enhances the accuracy and reliability of medical image segmentation by incorporating complementary information, ultimately aiding clinicians in making more effective diagnosis.

## 8. Conclusion

ViT-based image segmentation techniques have shown predominant performance in a number of image-related applications, including medical images. The self-attention mechanism in ViTs enables the model to learn global relationships in the images. This paper presents a detailed discussion of several architectural modifications along with some latest trends and training techniques utilized to improve the performance of ViT-based medical image segmentation approaches.

Despite ViTs' limitation in capturing local correlations in medical images, researchers have introduced approaches combining ViTs and CNNs to capture both local and global perspectives efficiently. We also discuss HVT-based medical image segmentation techniques in detail and categorize them based on their placement in encoder-decoder-based medical image segmentation architectures. In addition, we also provide an organ-wise overview of how these ViTs and HVTs are used in various real-world medical image segmentation applications, showcasing their impact on diagnosis, treatment, and disease monitoring.

Looking ahead, the field of ViT-based medical image segmentation continues to evolve rapidly. While promising achievements have been made, open questions remain regarding data efficiency, explainability, and adaptation to specific clinical tasks. Addressing these challenges and exploring synergies with other cutting-edge techniques holds immense potential for further revolutionizing medical image analysis and ultimately improving patient care.